\documentclass[journal=jacsat,manuscript=article]{achemso}

\usepackage[version=3]{mhchem} 
\usepackage{xcolor}
\usepackage{amsfonts} 
\setkeys{acs}{keywords = true} 

\author{Vishal Singh Pawak}
\affiliation[First University]
{Department of Chemical Engineering, Indian Institute of Technology Ropar, Rupnagar}
\author{Vijay A. Loganathan}
\affiliation[Second University]
{Department of Civil Engineering, Indian Institute of Technology Ropar, Rupnagar}
\author{Manigandan Sabapathy}
\email{mani@iitrpr.ac.in}
\affiliation[First University]
{Department of Chemical Engineering, Indian Institute of Technology Ropar, Rupnagar}

\title[An \textsf{achemso} demo]
  {Efficient removal of nanoplastics from synthetic wastewater using electrocoagulation}

\abbreviations{IR,NMR,UV}

\keywords{Nanoplastics, electrocoagulation, polystyrene, nanoparticles, wastewater}

\begin{document}






\begin{abstract}
Nanoplastics are emerging contaminants that have now transformed into a worldwide environmental concern. It is a lesser-known fact that several emerging contaminants, such as bisphenol and perfluoro alkylates adsorbing on micro and nanoplastics, could invade the food chain and cause irreversible damage to human health and the environment. Even though wastewater treatment plants (WWTPs) have been around for a long time, their removal strategy needs to be improved since this is one of the main routes that micro and nanoplastics get into the environment. UV deterioration, mechanical stresses, and biological processes cause plastics to break apart and turn into smaller pieces. They get small enough to be called nanoplastics, i.e. $<$ 1 $\mu$m. We studied the removal of nanoplastics from synthetic wastewater using an electrocoagulation process. Although electrocoagulation (EC) is a well-established wastewater treatment process for microplastic removal, the research on nanoplastics is still in its infancy. We used the polystyrene nanoparticles as nanoplastics synthesized from the expanded polystyrene waste (EPS). For studies on electrocoagulation, aluminium electrodes were used in parallel combination at low voltage conditions. We take advantage of the release of gas bubbles from the process to enable the removal from the top by scraping them off. We have studied the influence of various process parameters on removing nanoplastics, such as electrode spacing, salt concentration, and applied voltage. In this study, we have found that a maximum removal efficiency of more than 95\% could be achieved at a specific electrolyte concentration and at pH 7.2 ± 0.3, which illustrates that electrocoagulation is a successful technique for removing nanoplastic pollutants from the aquatic environment. The advantage of the proposed method is that when nanoplastics and coagulants are mixed, they help make a foamy layer on top of the reactor that can be easily scraped off. Alternatively, the target pollutants can also be lifted off the interface using a dip coater which can then be automated to perform the task continuously on a cyclic basis. The results of this study could serve as baseline information for achieving massive nanoplastics cleanup on a larger scale in an eco-friendly way.
  \end{abstract}

\section{Introduction}
Industrial wastewater has become one of the significant sources of plastic pollution. Plastic is used in a wide range of products, and its production has increased exponentially in the last few decades \cite{pico2022microplastics}. The production
of plastic has risen rapidly since the 1960s. In 1950, only 2.3 billion metric tons of plastic were produced worldwide. As of 2019, global plastic production reached more than 369 million tons annually. It is estimated that it may exceed a billion tons by 2050 \cite{shrivastava2020strengthening}. Researchers around the globe envisage that the oceans will be filled with more plastic than fish. It is because the amount of plastics entering the sea is way higher than usual. Consequently, per the prediction, 99 per cent of seabirds would have ingested plastic. Among several industries, the waste generated through personal care and cosmetic industries causes potential threats to the land and aquatics, wherein 93\% of the wastes are linked to polyethene-derived beads \cite{gouin2015use}. These plastic wastes are emerging contaminants that have become a worldwide environmental concern. The fragmentation of these plastic wastes transforms them into tinier pieces. This process converts them into the nanoplastics of size ranging from 1 nm to 1000 nm over some time owing to the combined effect of UV degradation, mechanical stresses, and biological processes \cite{napper2019environmental,dawson2018turning,geyer2017production,bianco2020degradation,bond2018occurrence}. 

The emergence of nanoplastics in pollution has become a significant challenge because of their unique properties imparted by various hydrocarbons and functional groups. It is challenging to monitor owing to its size, leading to a lack of knowledge and understanding of their potential impacts\cite{lindeque2020we,poulain2018small,wayman2021fate}. Recently, few researchers have found the presence of nanoplastics at the North Pole, and the South Pole of the earth 
\cite{materic2022nanoplastics}. Unlike larger-size impurities, these nanoplastics do not end up in the stomach when inadvertently consumed by humans. They are carried deep into the lungs and eventually enter the human circulatory system by penetrating the cell-blood barrier \cite{materic2022nanoplastics}. Further, the ingestion of nanoplastics causes damage to vital organs, tissues and cells. Their tiny size enables them to pass into the cell membranes \cite{da2016nano,gaylarde2021nanoplastics,kundu2021identification,lehner2019emergence,revel2018micro}. The growing concern in recent times is that nanoplastics may act as carriers for toxic chemicals and pathogens due to its association via hydrophobic backbone and functional groups. A particular type of nanoplastics can also bind to toxins and other harmful 
substances, which can later enter the body. The toxins generated in this way may lead to various adverse health effects, including immune dysfunction, endocrine disruption, cancer, damage to the gastrointestinal tract, and kidney functions\cite{ramasamy2021review,sana2020effects}. 
 
Polystyrene is a plastic found in various forms, such as microplastics and nanoplastics. Thermocol sheets are foams 
made of polystyrene (PS), which is very light in weight ( 2\% PS and 98\% air) but strong enough to withstand enormous amounts of pressure and are also known as expanded polystyrene (EPS). Thermocol sheets are used for packaging fragile items and food, thermal insulation of buildings, and industrial operations (such
as insulation material) as they are effortless to use and 
inexpensive\cite{rahimi2017chemical}. The United States of America generated $\approx$2 million tonnes of polystyrene waste in 2012, of which only 0.9\% was retrieved through recycling. The EPS waste can be converted into valuable products using proper technology and sustainable processes that may bring down the environmental issues caused by such materials. This study shows the conversion of thermocol sheets into polystyrene nanoparticles, which were then used as representative nanoplastic contaminants in aqueous systems. Further, we examined the removal of these nanoplastics using the electrocoagulation process under various experimental conditions.

Electrocoagulation is a well-known process for separating pollutants from wastewater using coagulants generated in situ and air bubbles generated during the electrolysis. This electrolysis-based process was first used in mining and then in water, and sewage treatment plants \cite{bhaskar1984electroflotation}.
In this process, bubbles form on the electrodes (anode and cathode) and dissolve in the water, colliding with solid or liquid particles/ pollutants and bringing them to the surface. Electrocoagulation can be a fruitful technique to use instead of chemical treatment processes. Chemical treatments require the addition of coagulants and flocculants into the wastewater, which can be expensive and result in large volumes of non-recoverable sludge. On the other hand, electrocoagulation does not require the use of chemicals and could serve as a green and clean technique for wastewater treatment. 

In this work, we demonstrate the removal of nanoplastics by employing an electrocoagulation method with improved process parameters. In contrast to microplastics, the electrocoagulation-induced separation of nanoplastics involves waste removal from the top. The action of bubbles combined with coagulants such as Al$^{3+}$ released by an aluminium electrode (Anode) provides a required thrust so that the nanoplastics are pushed towards the interface. Intriguingly, the accumulating nanoparticles create a dense and transparent foam-like layer at the top, enabling us to scrape them off quickly. Since the aggregates generated are of a smaller size, they resist gravity. Hence, the process starts with creaming rather than sedimentation. In this way, the removal of nanoparticles with a separation efficiency of 95.4\% efficiency can be achieved using the proposed method, which is noteworthy. To the best of our knowledge, no or very few studies have been reported on the removal of nanoplastics concerning electrocoagulation.

\section{Materials and Methods}

\subsection{Materials}

We sourced the expanded polystyrene (EPS) from the remnants of the packaging materials. The analytical grade acetone, used as the solvent to initiate the precipitation process, was procured from Rankem, India. Further, uniform nanoplastic particles were synthesized, and their removal rate was measured using a turbidity meter. We used the electrodes of aluminium (Al) as an anode and cathode during the electrocoagulation process. To demonstrate the elimination of nanoplastics under salt conditions, we prepared an aqueous electrolyte solution using analytical-grade monovalent salt (NaCl) from Sigma-Aldrich, India. The deionized water (18.2 $M\Omega$) from the water purification system procured from Thermo Fisher Scientific India Pvt. Ltd (Smart2Pure$^{TM}$) was used for all aqueous-based investigations.

\subsection{Synthesis of PS particles from waste thermocol sheets}

\begin{figure}
    \centering
\includegraphics[width=1.0\textwidth]{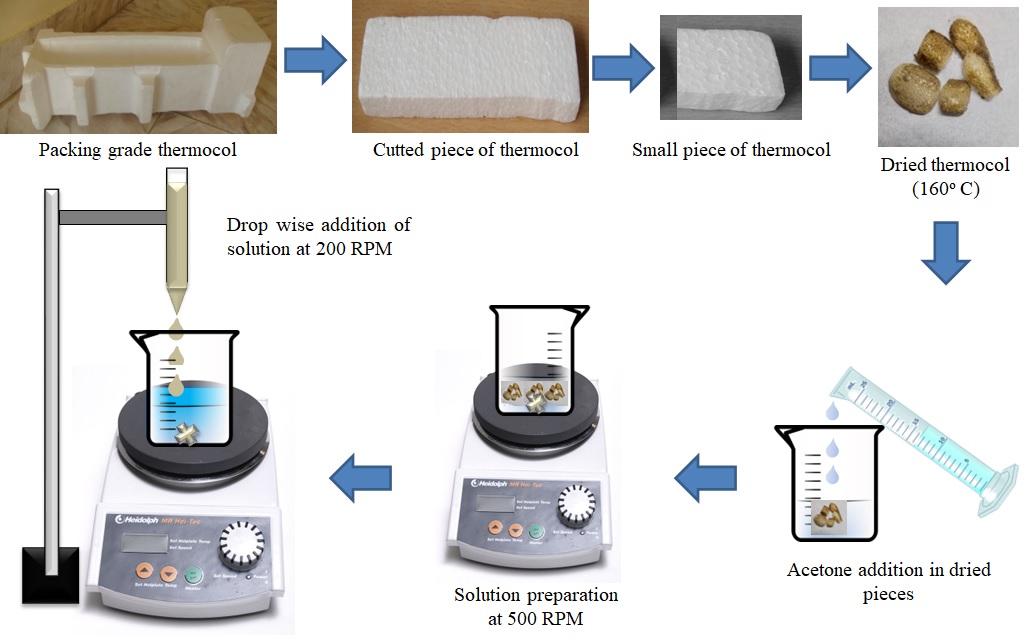}
      \caption{Schematic representation depicting the process of nanoparticle synthesis using EPS waste.}
    \label{scheme}
\end{figure}

To synthesize EPS particles using the nanoprecipitation technique, we followed the method described by \citeauthor {rajeev2016conversion} \cite{rajeev2016conversion} Firstly, packaging-grade thermocol was cut into smaller pieces and placed in a hot air oven at 160 $^{\circ}$C for 12 hours to remove volatile matter. This process caused the thermocol pieces to shrink, as depicted in Figure \ref{scheme}, which illustrates the step-by-step process of the nanoprecipitation technique followed to precipitate PS nanoparticles. Subsequently, we dissolved the dried samples in acetone at a 1:20 (EPS: acetone) ratio by mass and stirred the mixture using a magnetic stirrer at 500 rpm for 6 h. We allowed impurities to settle and separated the clear solution into a container. For nanoparticle synthesis, 1 mL of EPS solution made of acetone was continuously poured using a burette into a beaker containing 20 mL of deionized water. The drop-wise addition of solution was done at a flow rate of 0.5 mL/minute under constant stirring at a speed of 200 rpm and at room temperature. As a result, water present in the beaker instantly becomes turbid, signifying the precipitation of PS nanoparticles. The suspension containing acetone and EPS was kept at 25 $^{\circ}$C and 60 $^{\circ}$C for 8 and 4 h, respectively, to evaporate the acetone and excess solvent. Note: We have given sufficient time for precipitation before heating the mixture to 60 $^{\circ}$C for 4 h to achieve uniform distribution. Subsequently, we characterized the particles using several characterization techniques. The average diameter and zeta potential of the as-synthesized EPS particles measured using dynamic light scattering and electrophoretic light scattering were 175.6 $\pm$ 0.83 nm and -39 $\pm$ 0.4 mV, respectively.

\subsection{Electrocoagulation}

The experiments were conducted using synthetic wastewater (100 mL) and nanoplastics of size 175.6 ± 0.83 nm. The concentration of nanoplastics used in suspension was 0.35 mg/mL. The concentration range chosen is in agreement with several studies reported in literature \cite{perren2018removal,shen2022efficient}. The studies were carried out in a batch 
reactor with aluminium electrodes, in parallel combination, with a continuous stirring at a speed of 150 rpm \cite{shen2022efficient}. We have used Keithley's (triple channel) DC power supply to apply the appropriate voltage, as shown in figure \ref{EC} A. The effects of the various experimental parameters, such as electrode spacing, salt (NaCl) 
concentration and applied voltage, on removal efficiency were examined. During this process, two aluminium electrodes of dimensions of 90 mm x 12 mm x 2 mm each were used with an electrode spacing of 1 cm and 2 cm. The concentration of NaCl was varied (i.e. 10, 30 and 50 mM) to investigate the influence of varying current intensities on the removal of nanoplastics. The effect of applied voltage density on nanoplastics removal was studied under 5 V, 10 V and 15 V scenarios at a constant pH of 7.2. For each experiment, the electrochemical set-up was run for 3, 5, 10, 15, and 20 minutes before being switched off. The contaminants were allowed to get close to the interface and stay on top for up to 5 minutes. Subsequently, the nanoplastics generated near the interface turn turbid or foamy, leaving the solution clear. After each experiment, electrodes were washed in an acid bath of 1 M Hydrochloric acid (HCl) for at least 30 minutes before reuse.

\subsection{Removal efficiency measurement}

We have used the turbidity-based technique to quantify the removal of nanoplastics from synthetic wastewater as defined by \citeauthor {gregory1998turbidity} \cite{gregory1998turbidity}.
Turbidity measures the degree of suspended particles in water and is typically measured in NTU (Nephelometric Turbidity Units). It is an essential parameter in water quality monitoring as it can indicate the presence of pollutants and other contaminants \cite{gregory1998turbidity}. Turbidity has been employed recently in several studies with good selectivity and found to be an effective technique for quantifying 
nanoparticles \cite{wyatt2018measuring}. Turbidity measurement was used by \citeauthor {arenas2021nanoplastics} to calculate the removal efficiency of nanoplastics by GAC in a complex matrix, and they concluded that it is a valuable and efficient method for determining 
nanoplastic concentrations in wastewater \cite{arenas2021nanoplastics}. In this work, we have used the same approach to determine the removal efficiency of 
nanoplastics before and after the electrocoagulation process.

The removal efficiency of nanoplastics (R\%) for each experiment was measured using
the Eq. \ref{eq.1} below.

\begin{equation}
R \% = \frac{C_i-C_f}{C_i}\times100
\label{eq.1}
\end{equation}
                   
Where ${C_i}$ and ${C_f}$ represent the initial and final turbidity of the 
wastewater before and after the electrocoagulation process, respectively.

\subsection{Characterization}

The as-synthesized nanoplastics were analyzed using a field emission scanning electron microscopy (FESEM), Make: FEI, USA and Model: Inspect F50, to characterize the particle morphology and deduce the average size. The hydrodynamic radius of EPS particles was inferred using the dynamic light scattering (DLS), Make: Malvern Instruments and Model: Zetasizer Nano ZSP. The zeta potential of EPS was measured based on the electrophoretic light scattering (ELS) study using Zetasizer procured from Malvern Instruments, Model: Zetasizer Nano ZSP. The turbidity meter (Make: OAKTON, Model: T100) was employed to obtain nephelometric turbidity units (NTU) as a function of the time of electrocoagulation to investigate the creaming process. Fourier-transform Infrared spectroscopy (Make: Thermo Scientific Nicolet, Model: iS50) was used to identify the chemical composition of EPS particles. Thermal gravimetric analysis (TGA), make: TA instrument, model: SDT650, was carried out to capture the decomposition profiles of the EPS particles.

\begin{figure}
    \centering
\includegraphics[width=1.0\textwidth]{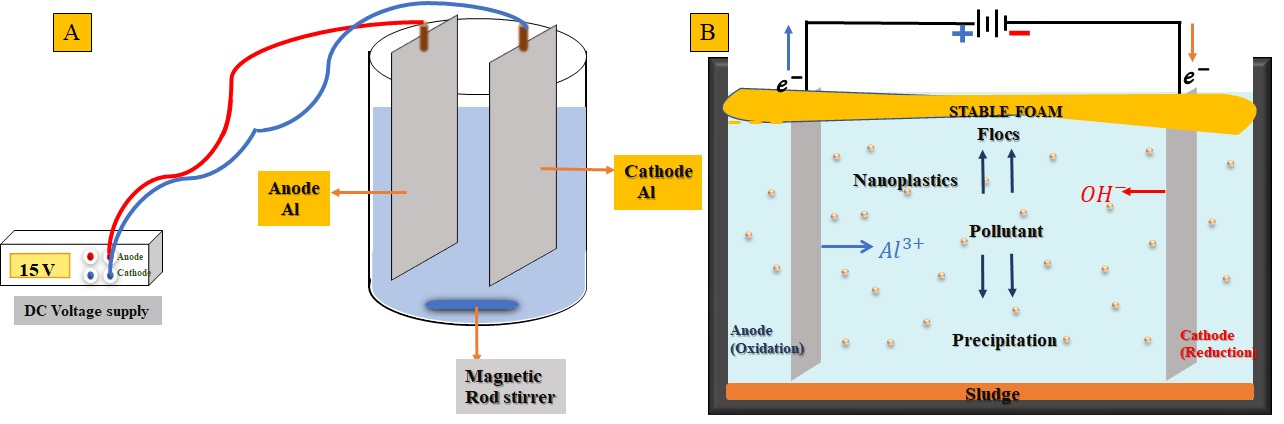}
        \caption{Schematic illustration showing (a) experimental setup (b) Mechanism of nanoplastics removal using Electrocoagulation process.}
    \label{EC}
\end{figure}

\section{Results and Discussion}

\begin{figure}
    \centering
\includegraphics[width=1.0\textwidth]{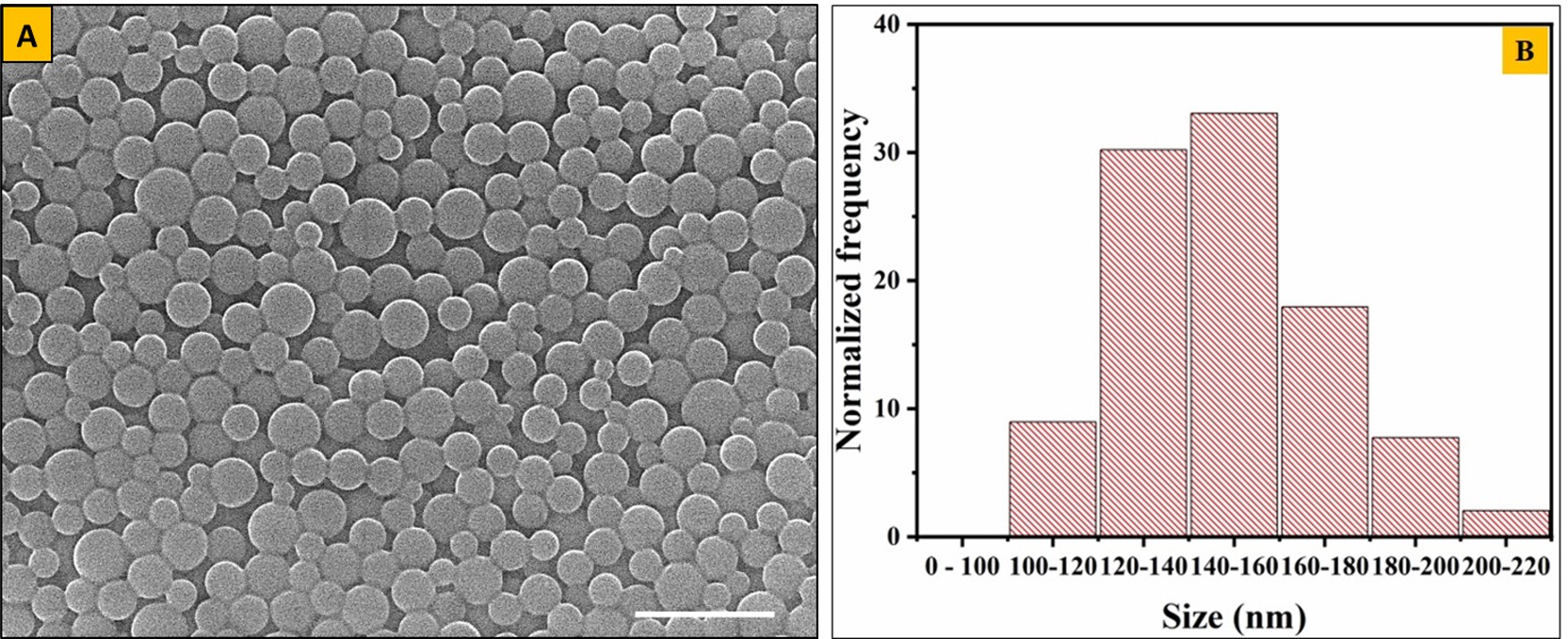}
    \caption{Structural characterization of EPS particles. A) FESEM image, B) size distribution obtained by analyzing the FESEM images using ImageJ. The scale bar given in the image corresponds to 500 nm.}
    \label{SEM}
\end{figure}

First, we discuss the characteristic features of EPS particles inferred from various characterization techniques. Figure \ref{SEM} A depicts the representative scanning electron microscopic image corresponding to the EPS surface morphology. The particles obtained are spherical and slightly poly-dispersed. Figure \ref{SEM} B depicts the size distribution of nanoplastics obtained by performing image analysis using ImageJ software. Based on 250 representative particles, the average size is 148 $\pm$ 22 nm. The close examination of the several reported data from the literature indicates that the deviation of 15-20\% is quite typical in synthesizing polystyrene (PS) nanoparticles from the EPS waste. Nevertheless, such variation is not detrimental to our studies considering the polydispersity associated with the microplastics and nanoplastics distribution in real-time application.  

\begin{figure}
    \centering
\includegraphics[width=1.0\textwidth]{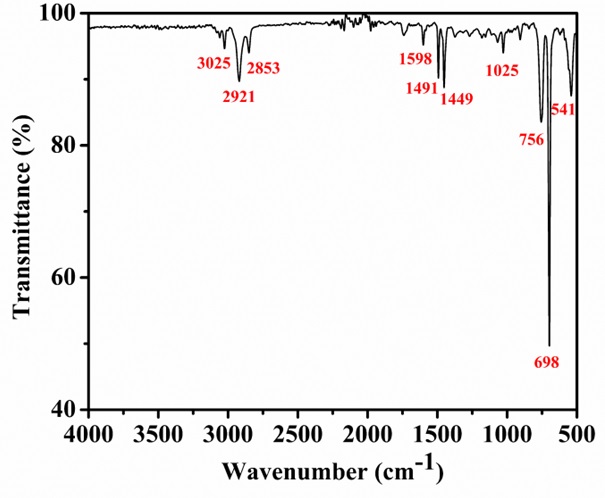}
       \caption{FTIR spectra of polystyrene nanoparticles made from EPS waste using the nanoprecipitation method.}
    \label{FTIR}
\end{figure}

Figure \ref{FTIR} displays the FTIR spectrum of as-synthesized EPS nanoparticles. The spectra observed between 3060-2853 $cm^{-1}$ correspond to the characteristic bands for aromatic and aliphatic C-H stretching, while the peak at 1598 $cm^{-1}$ reiterates the presence of aromatic stretch, i.e., C=C. On the other hand, the peak at 3443 $cm^{-1}$ indicates the existence of OH stretching. In short, the appearance of characteristic peaks at three different regions, 1) 400-900 cm$^{-1}$, 2) 900-2000 cm$^{-1}$, and 3) 2800-3200 cm$^{-1}$, reaffirm its existence identical to that of the chemical composition of PS. Further, the characteristics of polystyrene observed by us are in good agreement with the FTIR spectrum reported by \citeauthor {rajeev2016conversion} \cite{rajeev2016conversion}

\begin{figure}
    \centering
\includegraphics[width=1.0\textwidth]{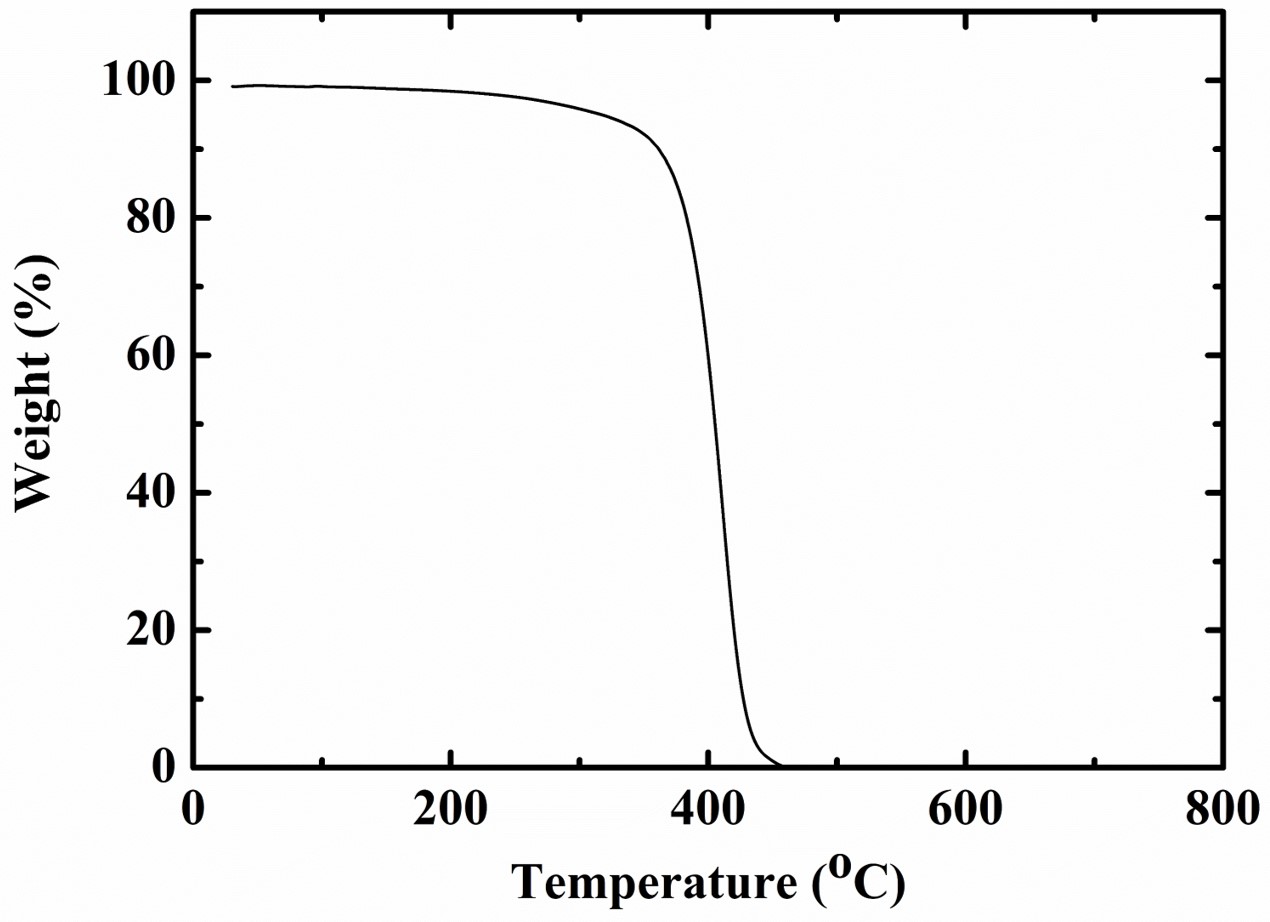}
       \caption{Thermal gravimetric analysis (TGA) of polystyrene nanoparticles obtained from EPS waste.}
    \label{TGA}
\end{figure}

Figure \ref{TGA} displays the thermal decomposition profile of polystyrene nanoparticles obtained from EPS waste. The respective polystyrene sample was subjected to temperature sweep from 30$^{\circ}$C to 800$^{\circ}$C with a heating rate of 10$^{\circ}$C/min. The decomposition data vis-a-vis temperature was in good agreement with the reported values of polystyrene, i.e., initial decomposition temperature - 300$^{\circ}$C and corresponding half decomposition temperature - 364$^{\circ}$C.\cite{rajeev2016conversion}   

Thus, combining various characterization techniques discussed so far establishes the physicochemical properties close to that of pure polystyrene materials. Therefore, the as-synthesized nanoparticles can be treated similarly to the nanoplastics, which is paramount in pursuing the electrocoagulation-based wastewater treatment containing plastic particles. The basis for choosing negatively charged polystyrene instead of positively charged one is that almost 93\% of micro or nanoplastics waste collected from various channels are polyethene/polyethylene derived compounds,  which are negatively charged. \cite{gouin2015use} Hence, to simulate similar conditions, we pursued our study using the negatively charged polystyrene particles less than 1 $\mu$m by recovering them from the EPS waste.

Before heading the topic into the removal of nanoplastics, we discuss the underlying principles of the proposed method induced by electrocoagulation. In this process, coagulants are generated in situ due to the oxidation of the metal anode when an electric current is passed through it \cite{pulkka2014electrochemical}. At the anode, $Al^{3+}$ ions are liberated and dissolved in the suspension. The dissolved metal ions combine with hydroxyl ions in the water to form metal hydroxides acting as coagulants \cite{behbahani2011comparison}. These coagulants neutralize the surface charges of pollutants/nanoplastics, leading to 
the movement of particles closer to each other due to the decrease in the electrostatic force of repulsion. It is believed that the $Al(OH)_3$ produced by anodic Al dissolution is more effective at coagulating pollutants in wastewater. However, the passivation of Al anodes and the impermeable film formed on cathodes can hinder the performance of electrocoagulation \cite{holt2005future}. Following reactions are carried out at the anode and cathode, \cite{perren2018removal,shen2022efficient}.

At Anode:
\begin{equation}
    Al_{(s)} \longrightarrow Al_{(aq)} {^{3+}} + 3e^-
    \label{eq.2}
    \end{equation}
\begin{equation}
    2 {H_2}O_{(l)} \longrightarrow 4 H_{(aq)} {^{+}} + {O_2}_{(g)} + 
    4e^-
    \label{eq.3}
\end{equation}

At Cathode:
\begin{equation}
    Al_{(aq)} {^{3+}} + 3e^- \longrightarrow Al_{(s)}
    \label{eq.4}
\end{equation}
\begin{equation}
    2 {H_2}O_{(l)} + 2 e^- \longrightarrow {H_2}_{(g)} + 2 OH{^{-}}
    \label{eq.5}
\end{equation}
\begin{equation}
    Al_{(aq)} {^{3+}} + 3 OH{^{-}} \longrightarrow {Al(OH)_3}_{(s)}
    \label{eq.6}
\end{equation}

The bubbles of hydrogen and oxygen generated during the electrolysis of water move upwards in the liquid and lead to the effective removal of contaminants \cite{linares2009influence,arroyo2009effect,el2009assessment}. We believe the generated metal ions combine with the nanoplastics present in the water, which gets removed via the formation of foam collected on the top of the reactor. This 
foam formation depends on several factors, such as hydrogen and oxygen gas bubble formation, type of contaminant, surface tension, zeta potential, temperature, and pH of the system for effective interaction of the gas bubbles and particles in the wastewater. From the eq. \ref{eq.3} and \ref{eq.5} mentioned above, the liberation of oxygen (${O_2}$) and hydrogen (${H_2}$) gas takes place at the anode and cathode, respectively, that forms gas bubbles (please refer to Figure \ref{EC}B). The real-time video capturing the electrolysis with the release of gas bubbles at respective electrodes can be found in the supplementary information section (SI). These gas bubbles will entrap many organic compounds and pollutants in the wastewater as they rise. All these particles entrapped in the gas bubbles rise and get collected in the vicinity of the interface, causing foam formation, which may be skimmed off later to remove these compounds from the water.

To this end, we show the removal of as-synthesized nanoplastics using the electrocoagulation process, wherein treatment of these nanoplastics was dependent on various operating parameters such as inter-electrode distance, applied 
voltage, and electrolyte concentration. We investigated the effect of these parameters to improve the removal rate of nanoplastics from wastewater. We also demonstrate the formed foam's stability after removing pollutants. 
However, through control studies, we initially examined the pollutant's removal and settling without the application of electrocoagulation. We observed the evidence of the removal of particles in the presence of electrolytes due to charge neutralization and accumulation of nanoplastics. On the other hand, no significant particle removal was observed in the absence of electrolytes. The removal efficiency in the absence of electrolytes was only 1.25\%, while it was 99.6\% in the presence of electrolytes after 27 days.

\subsection{Effect of electrode spacing}

To understand the effect of electrode spacing, we have varied the distance between the aluminium electrodes from  1 cm to 2 cm while keeping other parameters constant. According to many investigations, inter-electrode spacing has played an essential role in electrocoagulation. According to \citeauthor{bukhari2008investigation}, increasing the distance between the electrodes may decrease its efficiency and hence can reduce the capital cost for removing total suspended solids (TSS) from municipal waste \cite{bukhari2008investigation}. 

\begin{figure}
    \centering
    \includegraphics[width=1.0\textwidth]{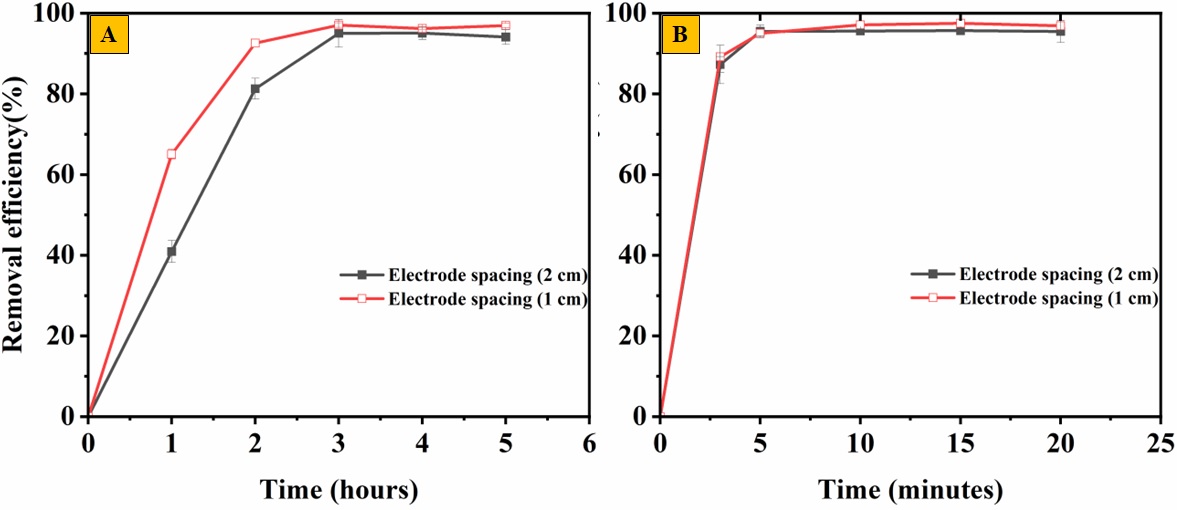}
        \caption{(a) Effect of electrode spacing on the removal of nanoplastics from wastewater in the electrocoagulation process in the absence of electrolyte. (b) Effect of electrode spacing with electrolyte (NaCl) on the removal of nanoplastics from wastewater using electrocoagulation process.}
    \label{fig6}
\end{figure}

The effect of electrode spacing on the removal efficiency is shown in Figure \ref{fig6}A. The experiments corresponding to the one shown in Figure \ref{fig6}A were performed without electrolyte. The other experimental conditions are as follows: Concentration of nanoplastics - 0.35 mg/ml, inter-electrode distance - 1 \& 2 cm, stirring speed - 150 rpm, applied voltage - 15 V, pH of the solution - 7.2. The removal rate found were 65.7 \%, 91.5 \%, 93 \%, 95.1 \%, 93.7 \%  at 1 cm and 40.8 \%, 80.1 \%, 91.2 \%, 93.3 \%, 92.2 \% at 2 cm inter-electrode distance after 1, 2, 3, 4 and 5 hours of electrocoagulation process, respectively, with electrodes being the aluminium. However, the removal of nanoplastics was significantly improved in the initial 3 hours by decreasing the electrode spacing from 2 cm to 1 cm. This enhanced performance can be attributed to the electrostatic effects, which could cause higher movement of ions, leading to a high creaming rate. \cite{khandegar2014electrochemical}.

The impact of nanoplastic removal was studied with varying inter-electrode distances at a fixed electrolyte (NaCl) concentration of 50 mM. As shown in Figure \ref{fig6}B, the removal rate of nanoplastics was greatly affected by the 
addition of electrolytes as it reduced the total time of the removal rate from 3 hours to 5 minutes. Although varying the distance between the electrodes did not show any significant effect in the presence of an electrolyte, it showed a slight increment in the removal efficiency of nanoplastics when the electrode distance was reduced from 2 cm to 1 cm. Other operating parameters are as follows: Concentration of nanoplastics - 0.35 mg/mL, pH of the solution - 7.2, electrolyte concentration - 50 mM, stirring speed - 150 rpm, and the applied voltage - 15 V. The removal efficiency of nanoplastics achieved was 87.2 \%, 95.4 \%, 95.5\%, 95.6 \%, 93.33 \% and 89.1 \%, 94.9 \%, 97.1 \%, 97.4 \%, 96.8 \%  at 2 cm and 1 cm electrode spacing after 3, 5, 10, 15, and 20 minutes of electrocoagulation process, respectively. The addition of electrolyte in the solution increases the dissolution of $Al^{3+}$ ions. Consequently, significant screening of nanoplastics' charge was achieved primarily due to these ions \cite{shen2022efficient}. 
 
\subsection{Effect of applied voltage}

The effect of voltage on the electrocoagulation process is paramount to improving the reaction rate of the process. According to Faraday's law, the amount of dissolution of $Al^{3+}$ ions increases when the amount of current passing through the Al anode increases. These $Al^{3+}$ ions dissolve into the solution and combine with the $OH^{-}$ present in water to form the aluminium hydroxide, which helps in increasing the removal efficiency of the nanoplastics \cite{shen2022efficient}.
Further, the increased applied voltage also increases the number density of hydrogen bubbles produced at the cathode. These hydrogen bubbles enhance the mixing phenomena of aluminium hydroxides with nanoplastics and improve the flotation of nanoplastics, which leads to an increase in the removal efficiency  \cite{guo2006enhanced,el2013treatment}. Furthermore, it was also found that by increasing the applied voltage, the amount of hydrogen bubbles increased, along with a reduction in the size, resulting in faster removal of nanoplastics and sludge flotation \cite{golder2007removal,el2013treatment}. 

The effect of applied voltage density on nanoplastic removal was performed under 5 V, 10 V and 15 V, respectively, while keeping the other operating parameters constant. Figure \ref{fig7}A shows the removal rate of nanoplastic over time under different applied voltage densities in the electrocoagulation process. The removal efficiency was observed 18.7 \%, 78.6 \%, 95.9 \%, 96.6 \%,95.8 \% at 5 V, 81.5 \%, 90.15 \%, 84.6 \%, 82.1 \%,77.0 \% at 10 V and 87.2 \%, 95.4 \%, 95.5 \%, 95.6 \%, 93.3 \% at 15 V, respectively, after 3, 5, 10, 15, 20 minutes of the electrocoagulation process.

\subsection{Effect of electrolyte concentrations}

\begin{figure}
    \centering
    \includegraphics[width=1.0\textwidth]{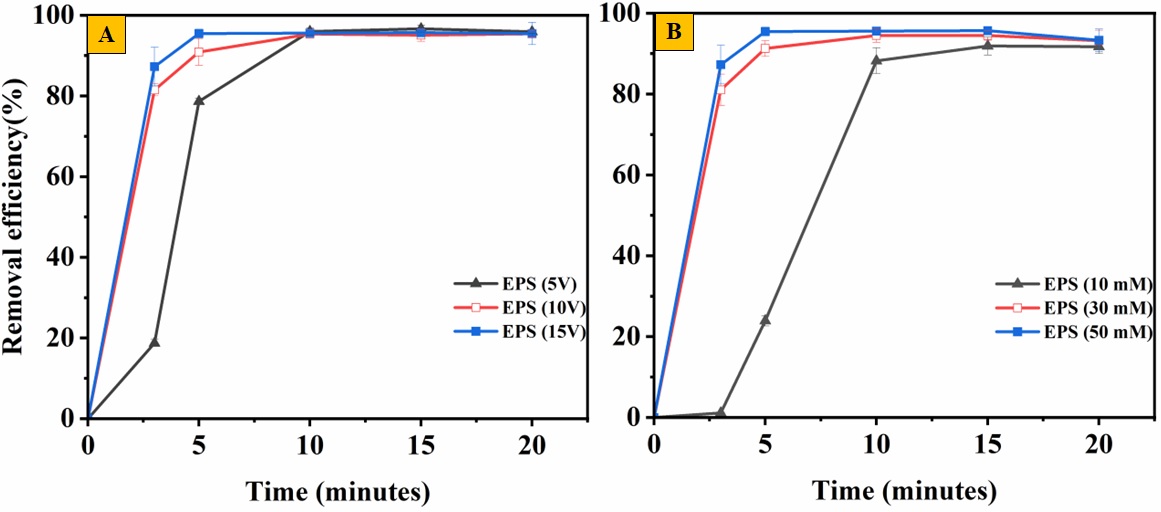}
        \caption{(a) Effect of applied voltage on the removal of nanoplastics from wastewater using electrocoagulation process (b) Effect of electrolyte concentration(NaCl) on the removal of nanoplastics from wastewater using electrocoagulation process.}
    \label{fig7}
\end{figure}

We have examined the effects of electrolyte concentration (NaCl) based on synthetic wastewater treatment. The electrolyte is a crucial parameter that reduces the reaction time and affects the operating cost and electricity consumption. The removal efficiency of nanoplastics was thoroughly investigated by varying the salt concentrations (NaCl) from 10mM to 50mM, as shown in Figure \ref{fig7} B. Adding electrolytes to the solution would increase the conductivity of the solution, thereby increasing the removal efficiency of microplastics\cite{shen2022efficient,perren2018removal}. \citeauthor{shalaby2014phosphate} demonstrated that the formation of $Al(OH)_3$ in the solution could be increased by adding the electrolytes. As a result, the process led to an enhanced phosphorus removal efficiency \cite{shalaby2014phosphate}. Our results show that the removal efficiency of nanoplastics was increased with an 
increase in electrolyte concentration from 10 mM to 50 mM with aluminium electrodes. We kept other parameters constant; concentration of nanoplastics - 0.35 mg/mL, electrode spacing - 2 cm, pH of the solution - 7.2, stirring speed - 150 rpm, and applied voltage - 15 V. The removal efficiency of nanoplastics achieved at different salt concentrations was 1.11 \%, 23.9 \%, 88.2 \%, 91.8 \%, 91.7 \% at 10 mM, 81.0 \%, 91.2 \%, 94.5 \%, 94.5, 93.1 \% at 30 mM and 87.2 \%, 95.4 \%,95.5 \%, 95.6 \%, 93.33 \% at 50 mM after 3, 5, 10, 15, and 20 minutes of the electrocoagulation process, respectively.

\subsection{Foam stability}

In electrocoagulation, foam stability significantly affects the process's efficiency and hence the treatment's effectiveness. Foam is generated during electrocoagulation due to the production of hydrogen and oxygen gases as bubbles at the electrodes. While these gas bubbles rise, they interact with the pollutants and drive them to the interface's vicinity. The collection of all nanoplastics generates a foamy layer-like appearance at the top.

Here, we denote the foam stability in terms of the removal efficiency of nanoplastics by measuring the turbidity of the clear phase at the bottom of the reactor. The idea is to monitor the NTU of bulk solution as a function of time to evaluate the net removal efficiency. If foam destabilizes over time, the nanoplastics trapped within the turbid phase will eventually diffuse back to the bulk. This phenomenon reverses the direction of the flow of the nanoparticles and thereby increases the turbidity in bulk. For deducing the net removal efficiency of the nanoplastics, we observed foam stability after running the reactor for 5 min at the optimized parameters. Note: We chose 5 min as a reference point as we observed the maximum efficiency within this time scale. After 5 minutes of operation, solid foam appeared on top of the container and turbidity of the clear phase was measured. We noticed that the foam stability started decreasing from 5 minutes to 95 minutes. Therefore we conclude that the turbidity of the solution induced by destabilization increases with time due to gravitational force, as shown in Figure \ref{EC}B.

\begin{figure}
    \centering
\includegraphics[width=1.0\textwidth]{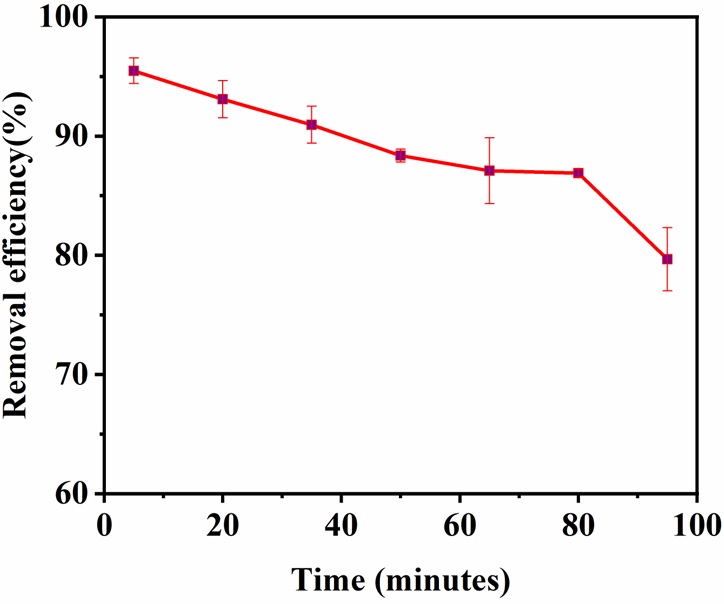}
       \caption{Stability analysis of the generated foam as a function of time}
    \label{stability}
\end{figure}

The removal efficiency of nanoplastics observed was 95.4 \%, 93 \%, 90.9\%, 88.3
\%, 87.0 \%, 86.8 \%, and 79.6 \%  after 5, 20, 35, 50, 65, 80, and 95 minutes, 
respectively, when the reactor was switched off after 5 minutes of 
electrocoagulation process.  As the foam got destabilized, the turbidity of the 
clear phase increased and the removal efficiency decreased over time 
(please refer to Figure \ref{stability}). We observed that the stability of the foam
decreased from 5 minutes to 95 minutes, with a maximal stability and removal 
efficiency of 95.45\% at 5 minutes and a minimum of 79.65\% at 95 minutes, 
indicating significant destabilization of the foam. Additionally, as time 
progressed, the foam became completely destabilized, and the clear phase 
became turbid, indicating that the solution is contaminated with  resuspension of pollutants in the synthetic water. It comes to say that it is essential to complete required scraping operation within certain duration to ensure maximum separation. For instance, one must perform necessary actions within 20 min to eliminate 93 \% of pollutants or within 35 min to achieve 91 \% removal of nanoplastics (pollutants). Therefore, the given stability data may be referred to as baseline information before performing a cleanup on a large scale.

\section{Conclusion}

This work demonstrates the electrocoagulation process as an effective method for removing nanoplastic waste from synthetic wastewater. The proposed process is an efficient technique with higher removal rates, lower energy consumption and sludge generation with low environmental impact and can be a green technique for nanoplastic removal. Further, the optimum parameters for removing synthesized PS nanoplastics were achieved at an electrolyte (NaCl) concentration of 50 mM, pH of the solution of 7.2, electrode spacing of 2 cm and an applied voltage of 15V. The removal of 95.4 \% of pollutants from synthetic wastewater containing nanoplastics at a concentration of 0.35 mg/mL or $\approx$ 350 ppm was achieved after 5 minutes of the electrocoagulation process. Eliminating nanoplastics using electrocoagulation is a practical and effective solution for addressing the growing problem of nanoplastics in the environment. Further research is needed to optimize the effectiveness of this process for different types of nanoplastics present in wastewater. However, electrocoagulation has the potential to play a crucial role in the management and reduction of nanoplastics pollution in the aquatic environment. It may be used as an effective and green method to treat larger volumes of wastewater with higher concentrations of nanoplastics.

\begin{acknowledgement}

This work was conducted through the support of Water and Soil Quality Assessment domain, AWaDH, IIT Ropar. The authors acknowledge the grant received from the Department of Science and Technology, Government of India, for the Technology Innovation Hub at the Indian Institute of Technology Ropar in the framework of National Mission on Interdisciplinary Cyber-Physical Systems (NM-ICPS). VSP acknowledges the guidance of Chandra Shekhar for synthesizing the polystyrene nanoparticles from EPS waste.

\end{acknowledgement}





\begin{mcitethebibliography}{42}
\providecommand*\natexlab[1]{#1}
\providecommand*\mciteSetBstSublistMode[1]{}
\providecommand*\mciteSetBstMaxWidthForm[2]{}
\providecommand*\mciteBstWouldAddEndPuncttrue
  {\def\EndOfBibitem{\unskip.}}
\providecommand*\mciteBstWouldAddEndPunctfalse
  {\let\EndOfBibitem\relax}
\providecommand*\mciteSetBstMidEndSepPunct[3]{}
\providecommand*\mciteSetBstSublistLabelBeginEnd[3]{}
\providecommand*\EndOfBibitem{}
\mciteSetBstSublistMode{f}
\mciteSetBstMaxWidthForm{subitem}{(\alph{mcitesubitemcount})}
\mciteSetBstSublistLabelBeginEnd
  {\mcitemaxwidthsubitemform\space}
  {\relax}
  {\relax}

\bibitem[Pic{\'o} \latin{et~al.}(2022)Pic{\'o}, Manzoor, Soursou, and
  Barcel{\'o}]{pico2022microplastics}
Pic{\'o},~Y.; Manzoor,~I.; Soursou,~V.; Barcel{\'o},~D. Microplastics in water,
  from treatment process to drinking water: analytical methods and potential
  health effects. \emph{Water Emerging Contaminants \& Nanoplastics}
  \textbf{2022}, \emph{1}, 13\relax
\mciteBstWouldAddEndPuncttrue
\mciteSetBstMidEndSepPunct{\mcitedefaultmidpunct}
{\mcitedefaultendpunct}{\mcitedefaultseppunct}\relax
\EndOfBibitem
\bibitem[Shrivastava \latin{et~al.}(2020)Shrivastava, Shrivastava,
  \latin{et~al.} others]{shrivastava2020strengthening}
Shrivastava,~S.~R.; Shrivastava,~P.~S., \latin{et~al.}  Strengthening of
  existing water treatment procedures to respond to the presence of
  microplastics in the drinking water. \emph{International Journal of
  Environmental Health Engineering} \textbf{2020}, \emph{9}, 1\relax
\mciteBstWouldAddEndPuncttrue
\mciteSetBstMidEndSepPunct{\mcitedefaultmidpunct}
{\mcitedefaultendpunct}{\mcitedefaultseppunct}\relax
\EndOfBibitem
\bibitem[Gouin \latin{et~al.}(2015)Gouin, Avalos, Brunning, Brzuska, De~Graaf,
  Kaumanns, Koning, Meyberg, Rettinger, Schlatter, \latin{et~al.}
  others]{gouin2015use}
Gouin,~T.; Avalos,~J.; Brunning,~I.; Brzuska,~K.; De~Graaf,~J.; Kaumanns,~J.;
  Koning,~T.; Meyberg,~M.; Rettinger,~K.; Schlatter,~H., \latin{et~al.}  Use of
  micro-plastic beads in cosmetic products in Europe and their estimated
  emissions to the North Sea environment. \emph{SOFW J} \textbf{2015},
  \emph{141}, 40--46\relax
\mciteBstWouldAddEndPuncttrue
\mciteSetBstMidEndSepPunct{\mcitedefaultmidpunct}
{\mcitedefaultendpunct}{\mcitedefaultseppunct}\relax
\EndOfBibitem
\bibitem[Napper and Thompson(2019)Napper, and
  Thompson]{napper2019environmental}
Napper,~I.~E.; Thompson,~R.~C. Environmental deterioration of biodegradable,
  oxo-biodegradable, compostable, and conventional plastic carrier bags in the
  sea, soil, and open-air over a 3-year period. \emph{Environmental science \&
  technology} \textbf{2019}, \emph{53}, 4775--4783\relax
\mciteBstWouldAddEndPuncttrue
\mciteSetBstMidEndSepPunct{\mcitedefaultmidpunct}
{\mcitedefaultendpunct}{\mcitedefaultseppunct}\relax
\EndOfBibitem
\bibitem[Dawson \latin{et~al.}(2018)Dawson, Kawaguchi, King, Townsend, King,
  Huston, and Bengtson~Nash]{dawson2018turning}
Dawson,~A.~L.; Kawaguchi,~S.; King,~C.~K.; Townsend,~K.~A.; King,~R.;
  Huston,~W.~M.; Bengtson~Nash,~S.~M. Turning microplastics into nanoplastics
  through digestive fragmentation by Antarctic krill. \emph{Nature
  communications} \textbf{2018}, \emph{9}, 1--8\relax
\mciteBstWouldAddEndPuncttrue
\mciteSetBstMidEndSepPunct{\mcitedefaultmidpunct}
{\mcitedefaultendpunct}{\mcitedefaultseppunct}\relax
\EndOfBibitem
\bibitem[Geyer \latin{et~al.}(2017)Geyer, Jambeck, and
  Law]{geyer2017production}
Geyer,~R.; Jambeck,~J.~R.; Law,~K.~L. Production, use, and fate of all plastics
  ever made. \emph{Science advances} \textbf{2017}, \emph{3}, e1700782\relax
\mciteBstWouldAddEndPuncttrue
\mciteSetBstMidEndSepPunct{\mcitedefaultmidpunct}
{\mcitedefaultendpunct}{\mcitedefaultseppunct}\relax
\EndOfBibitem
\bibitem[Bianco \latin{et~al.}(2020)Bianco, Sordello, Ehn, Vione, and
  Passananti]{bianco2020degradation}
Bianco,~A.; Sordello,~F.; Ehn,~M.; Vione,~D.; Passananti,~M. Degradation of
  nanoplastics in the environment: Reactivity and impact on atmospheric and
  surface waters. \emph{Science of the Total Environment} \textbf{2020},
  \emph{742}, 140413\relax
\mciteBstWouldAddEndPuncttrue
\mciteSetBstMidEndSepPunct{\mcitedefaultmidpunct}
{\mcitedefaultendpunct}{\mcitedefaultseppunct}\relax
\EndOfBibitem
\bibitem[Bond \latin{et~al.}(2018)Bond, Ferrandiz-Mas, Felipe-Sotelo, and
  Van~Sebille]{bond2018occurrence}
Bond,~T.; Ferrandiz-Mas,~V.; Felipe-Sotelo,~M.; Van~Sebille,~E. The occurrence
  and degradation of aquatic plastic litter based on polymer physicochemical
  properties: A review. \emph{Critical reviews in environmental science and
  technology} \textbf{2018}, \emph{48}, 685--722\relax
\mciteBstWouldAddEndPuncttrue
\mciteSetBstMidEndSepPunct{\mcitedefaultmidpunct}
{\mcitedefaultendpunct}{\mcitedefaultseppunct}\relax
\EndOfBibitem
\bibitem[Lindeque \latin{et~al.}(2020)Lindeque, Cole, Coppock, Lewis, Miller,
  Watts, Wilson-McNeal, Wright, and Galloway]{lindeque2020we}
Lindeque,~P.~K.; Cole,~M.; Coppock,~R.~L.; Lewis,~C.~N.; Miller,~R.~Z.;
  Watts,~A.~J.; Wilson-McNeal,~A.; Wright,~S.~L.; Galloway,~T.~S. Are we
  underestimating microplastic abundance in the marine environment? A
  comparison of microplastic capture with nets of different mesh-size.
  \emph{Environmental Pollution} \textbf{2020}, \emph{265}, 114721\relax
\mciteBstWouldAddEndPuncttrue
\mciteSetBstMidEndSepPunct{\mcitedefaultmidpunct}
{\mcitedefaultendpunct}{\mcitedefaultseppunct}\relax
\EndOfBibitem
\bibitem[Poulain \latin{et~al.}(2018)Poulain, Mercier, Brach, Martignac,
  Routaboul, Perez, Desjean, and Ter~Halle]{poulain2018small}
Poulain,~M.; Mercier,~M.~J.; Brach,~L.; Martignac,~M.; Routaboul,~C.;
  Perez,~E.; Desjean,~M.~C.; Ter~Halle,~A. Small microplastics as a main
  contributor to plastic mass balance in the North Atlantic subtropical gyre.
  \emph{Environmental science \& technology} \textbf{2018}, \emph{53},
  1157--1164\relax
\mciteBstWouldAddEndPuncttrue
\mciteSetBstMidEndSepPunct{\mcitedefaultmidpunct}
{\mcitedefaultendpunct}{\mcitedefaultseppunct}\relax
\EndOfBibitem
\bibitem[Wayman and Niemann(2021)Wayman, and Niemann]{wayman2021fate}
Wayman,~C.; Niemann,~H. The fate of plastic in the ocean environment--a
  minireview. \emph{Environmental Science: Processes \& Impacts} \textbf{2021},
  \emph{23}, 198--212\relax
\mciteBstWouldAddEndPuncttrue
\mciteSetBstMidEndSepPunct{\mcitedefaultmidpunct}
{\mcitedefaultendpunct}{\mcitedefaultseppunct}\relax
\EndOfBibitem
\bibitem[Materi{\'c} \latin{et~al.}(2022)Materi{\'c}, Kj{\ae}r, Vallelonga,
  Tison, R{\"o}ckmann, and Holzinger]{materic2022nanoplastics}
Materi{\'c},~D.; Kj{\ae}r,~H.~A.; Vallelonga,~P.; Tison,~J.-L.;
  R{\"o}ckmann,~T.; Holzinger,~R. Nanoplastics measurements in Northern and
  Southern polar ice. \emph{Environmental research} \textbf{2022}, \emph{208},
  112741\relax
\mciteBstWouldAddEndPuncttrue
\mciteSetBstMidEndSepPunct{\mcitedefaultmidpunct}
{\mcitedefaultendpunct}{\mcitedefaultseppunct}\relax
\EndOfBibitem
\bibitem[da~Costa \latin{et~al.}(2016)da~Costa, Santos, Duarte, and
  Rocha-Santos]{da2016nano}
da~Costa,~J.~P.; Santos,~P.~S.; Duarte,~A.~C.; Rocha-Santos,~T. (Nano) plastics
  in the environment--sources, fates and effects. \emph{Science of the total
  environment} \textbf{2016}, \emph{566}, 15--26\relax
\mciteBstWouldAddEndPuncttrue
\mciteSetBstMidEndSepPunct{\mcitedefaultmidpunct}
{\mcitedefaultendpunct}{\mcitedefaultseppunct}\relax
\EndOfBibitem
\bibitem[Gaylarde \latin{et~al.}(2021)Gaylarde, Neto, and
  da~Fonseca]{gaylarde2021nanoplastics}
Gaylarde,~C.~C.; Neto,~J. A.~B.; da~Fonseca,~E.~M. Nanoplastics in aquatic
  systems-are they more hazardous than microplastics? \emph{Environmental
  Pollution} \textbf{2021}, \emph{272}, 115950\relax
\mciteBstWouldAddEndPuncttrue
\mciteSetBstMidEndSepPunct{\mcitedefaultmidpunct}
{\mcitedefaultendpunct}{\mcitedefaultseppunct}\relax
\EndOfBibitem
\bibitem[Kundu \latin{et~al.}(2021)Kundu, Shetti, Basu, Reddy, Nadagouda, and
  Aminabhavi]{kundu2021identification}
Kundu,~A.; Shetti,~N.~P.; Basu,~S.; Reddy,~K.~R.; Nadagouda,~M.~N.;
  Aminabhavi,~T.~M. Identification and removal of micro-and nano-plastics:
  Efficient and cost-effective methods. \emph{Chemical Engineering Journal}
  \textbf{2021}, \emph{421}, 129816\relax
\mciteBstWouldAddEndPuncttrue
\mciteSetBstMidEndSepPunct{\mcitedefaultmidpunct}
{\mcitedefaultendpunct}{\mcitedefaultseppunct}\relax
\EndOfBibitem
\bibitem[Lehner \latin{et~al.}(2019)Lehner, Weder, Petri-Fink, and
  Rothen-Rutishauser]{lehner2019emergence}
Lehner,~R.; Weder,~C.; Petri-Fink,~A.; Rothen-Rutishauser,~B. Emergence of
  nanoplastic in the environment and possible impact on human health.
  \emph{Environmental science \& technology} \textbf{2019}, \emph{53},
  1748--1765\relax
\mciteBstWouldAddEndPuncttrue
\mciteSetBstMidEndSepPunct{\mcitedefaultmidpunct}
{\mcitedefaultendpunct}{\mcitedefaultseppunct}\relax
\EndOfBibitem
\bibitem[Revel \latin{et~al.}(2018)Revel, Ch{\^a}tel, and
  Mouneyrac]{revel2018micro}
Revel,~M.; Ch{\^a}tel,~A.; Mouneyrac,~C. Micro (nano) plastics: A threat to
  human health? \emph{Current Opinion in Environmental Science \& Health}
  \textbf{2018}, \emph{1}, 17--23\relax
\mciteBstWouldAddEndPuncttrue
\mciteSetBstMidEndSepPunct{\mcitedefaultmidpunct}
{\mcitedefaultendpunct}{\mcitedefaultseppunct}\relax
\EndOfBibitem
\bibitem[Ramasamy and Palanisamy(2021)Ramasamy, and
  Palanisamy]{ramasamy2021review}
Ramasamy,~B. S.~S.; Palanisamy,~S. A review on occurrence, characteristics,
  toxicology and treatment of nanoplastic waste in the environment.
  \emph{Environmental Science and Pollution Research} \textbf{2021}, \emph{28},
  43258--43273\relax
\mciteBstWouldAddEndPuncttrue
\mciteSetBstMidEndSepPunct{\mcitedefaultmidpunct}
{\mcitedefaultendpunct}{\mcitedefaultseppunct}\relax
\EndOfBibitem
\bibitem[Sana \latin{et~al.}(2020)Sana, Dogiparthi, Gangadhar, Chakravorty, and
  Abhishek]{sana2020effects}
Sana,~S.~S.; Dogiparthi,~L.~K.; Gangadhar,~L.; Chakravorty,~A.; Abhishek,~N.
  Effects of microplastics and nanoplastics on marine environment and human
  health. \emph{Environmental Science and Pollution Research} \textbf{2020},
  \emph{27}, 44743--44756\relax
\mciteBstWouldAddEndPuncttrue
\mciteSetBstMidEndSepPunct{\mcitedefaultmidpunct}
{\mcitedefaultendpunct}{\mcitedefaultseppunct}\relax
\EndOfBibitem
\bibitem[Rahimi and Garc{\'\i}a(2017)Rahimi, and
  Garc{\'\i}a]{rahimi2017chemical}
Rahimi,~A.; Garc{\'\i}a,~J.~M. Chemical recycling of waste plastics for new
  materials production. \emph{Nature Reviews Chemistry} \textbf{2017},
  \emph{1}, 1--11\relax
\mciteBstWouldAddEndPuncttrue
\mciteSetBstMidEndSepPunct{\mcitedefaultmidpunct}
{\mcitedefaultendpunct}{\mcitedefaultseppunct}\relax
\EndOfBibitem
\bibitem[Bhaskar~Raju and Khangaonkar(1984)Bhaskar~Raju, and
  Khangaonkar]{bhaskar1984electroflotation}
Bhaskar~Raju,~G.; Khangaonkar,~P. Electroflotation-A critical review.
  \emph{Transactions of the Indian Institute of Metals} \textbf{1984},
  \emph{37}, 59--66\relax
\mciteBstWouldAddEndPuncttrue
\mciteSetBstMidEndSepPunct{\mcitedefaultmidpunct}
{\mcitedefaultendpunct}{\mcitedefaultseppunct}\relax
\EndOfBibitem
\bibitem[Rajeev \latin{et~al.}(2016)Rajeev, Erapalapati, Madhavan, and
  Basavaraj]{rajeev2016conversion}
Rajeev,~A.; Erapalapati,~V.; Madhavan,~N.; Basavaraj,~M.~G. Conversion of
  expanded polystyrene waste to nanoparticles via nanoprecipitation.
  \emph{Journal of Applied Polymer Science} \textbf{2016}, \emph{133}\relax
\mciteBstWouldAddEndPuncttrue
\mciteSetBstMidEndSepPunct{\mcitedefaultmidpunct}
{\mcitedefaultendpunct}{\mcitedefaultseppunct}\relax
\EndOfBibitem
\bibitem[Perren \latin{et~al.}(2018)Perren, Wojtasik, and
  Cai]{perren2018removal}
Perren,~W.; Wojtasik,~A.; Cai,~Q. Removal of microbeads from wastewater using
  electrocoagulation. \emph{ACS omega} \textbf{2018}, \emph{3},
  3357--3364\relax
\mciteBstWouldAddEndPuncttrue
\mciteSetBstMidEndSepPunct{\mcitedefaultmidpunct}
{\mcitedefaultendpunct}{\mcitedefaultseppunct}\relax
\EndOfBibitem
\bibitem[Shen \latin{et~al.}(2022)Shen, Zhang, Almatrafi, Hu, Zhou, Song, Zeng,
  and Zeng]{shen2022efficient}
Shen,~M.; Zhang,~Y.; Almatrafi,~E.; Hu,~T.; Zhou,~C.; Song,~B.; Zeng,~Z.;
  Zeng,~G. Efficient removal of microplastics from wastewater by an
  electrocoagulation process. \emph{Chemical Engineering Journal}
  \textbf{2022}, \emph{428}, 131161\relax
\mciteBstWouldAddEndPuncttrue
\mciteSetBstMidEndSepPunct{\mcitedefaultmidpunct}
{\mcitedefaultendpunct}{\mcitedefaultseppunct}\relax
\EndOfBibitem
\bibitem[Gregory(1998)]{gregory1998turbidity}
Gregory,~J. Turbidity and beyond. \emph{Filtration \& separation}
  \textbf{1998}, \emph{35}, 63--67\relax
\mciteBstWouldAddEndPuncttrue
\mciteSetBstMidEndSepPunct{\mcitedefaultmidpunct}
{\mcitedefaultendpunct}{\mcitedefaultseppunct}\relax
\EndOfBibitem
\bibitem[Wyatt(2018)]{wyatt2018measuring}
Wyatt,~P.~J. Measuring nanoparticles in the size range to 2000 nm.
  \emph{Journal of Nanoparticle Research} \textbf{2018}, \emph{20}, 1--18\relax
\mciteBstWouldAddEndPuncttrue
\mciteSetBstMidEndSepPunct{\mcitedefaultmidpunct}
{\mcitedefaultendpunct}{\mcitedefaultseppunct}\relax
\EndOfBibitem
\bibitem[Arenas \latin{et~al.}(2021)Arenas, Gentile, Zimmermann, and
  Stoll]{arenas2021nanoplastics}
Arenas,~L.~R.; Gentile,~S.~R.; Zimmermann,~S.; Stoll,~S. Nanoplastics
  adsorption and removal efficiency by granular activated carbon used in
  drinking water treatment process. \emph{Science of the Total Environment}
  \textbf{2021}, \emph{791}, 148175\relax
\mciteBstWouldAddEndPuncttrue
\mciteSetBstMidEndSepPunct{\mcitedefaultmidpunct}
{\mcitedefaultendpunct}{\mcitedefaultseppunct}\relax
\EndOfBibitem
\bibitem[Pulkka \latin{et~al.}(2014)Pulkka, Martikainen, Bhatnagar, and
  Sillanp{\"a}{\"a}]{pulkka2014electrochemical}
Pulkka,~S.; Martikainen,~M.; Bhatnagar,~A.; Sillanp{\"a}{\"a},~M.
  Electrochemical methods for the removal of anionic contaminants from water--a
  review. \emph{Separation and Purification Technology} \textbf{2014},
  \emph{132}, 252--271\relax
\mciteBstWouldAddEndPuncttrue
\mciteSetBstMidEndSepPunct{\mcitedefaultmidpunct}
{\mcitedefaultendpunct}{\mcitedefaultseppunct}\relax
\EndOfBibitem
\bibitem[Behbahani \latin{et~al.}(2011)Behbahani, ALAVI, and
  Arami]{behbahani2011comparison}
Behbahani,~M.; ALAVI,~M.~M.; Arami,~M. A comparison between aluminum and iron
  electrodes on removal of phosphate from aqueous solutions by
  electrocoagulation process. \textbf{2011}, \emph{5}, 403-412\relax
\mciteBstWouldAddEndPunctfalse
\mciteSetBstMidEndSepPunct{\mcitedefaultmidpunct}
{}{\mcitedefaultseppunct}\relax
\EndOfBibitem
\bibitem[Holt \latin{et~al.}(2005)Holt, Barton, and Mitchell]{holt2005future}
Holt,~P.~K.; Barton,~G.~W.; Mitchell,~C.~A. The future for electrocoagulation
  as a localised water treatment technology. \emph{Chemosphere} \textbf{2005},
  \emph{59}, 355--367\relax
\mciteBstWouldAddEndPuncttrue
\mciteSetBstMidEndSepPunct{\mcitedefaultmidpunct}
{\mcitedefaultendpunct}{\mcitedefaultseppunct}\relax
\EndOfBibitem
\bibitem[Linares-Hern{\'a}ndez \latin{et~al.}(2009)Linares-Hern{\'a}ndez,
  Barrera-D{\'\i}az, Roa-Morales, Bilyeu, and
  Ure{\~n}a-N{\'u}{\~n}ez]{linares2009influence}
Linares-Hern{\'a}ndez,~I.; Barrera-D{\'\i}az,~C.; Roa-Morales,~G.; Bilyeu,~B.;
  Ure{\~n}a-N{\'u}{\~n}ez,~F. Influence of the anodic material on
  electrocoagulation performance. \emph{Chemical engineering journal}
  \textbf{2009}, \emph{148}, 97--105\relax
\mciteBstWouldAddEndPuncttrue
\mciteSetBstMidEndSepPunct{\mcitedefaultmidpunct}
{\mcitedefaultendpunct}{\mcitedefaultseppunct}\relax
\EndOfBibitem
\bibitem[Arroyo \latin{et~al.}(2009)Arroyo, P{\'e}rez-Herranz, Montanes,
  Garc{\'\i}a-Ant{\'o}n, and Guinon]{arroyo2009effect}
Arroyo,~M.; P{\'e}rez-Herranz,~V.; Montanes,~M.; Garc{\'\i}a-Ant{\'o}n,~J.;
  Guinon,~J. Effect of pH and chloride concentration on the removal of
  hexavalent chromium in a batch electrocoagulation reactor. \emph{Journal of
  Hazardous Materials} \textbf{2009}, \emph{169}, 1127--1133\relax
\mciteBstWouldAddEndPuncttrue
\mciteSetBstMidEndSepPunct{\mcitedefaultmidpunct}
{\mcitedefaultendpunct}{\mcitedefaultseppunct}\relax
\EndOfBibitem
\bibitem[El-Naas \latin{et~al.}(2009)El-Naas, Al-Zuhair, Al-Lobaney, and
  Makhlouf]{el2009assessment}
El-Naas,~M.~H.; Al-Zuhair,~S.; Al-Lobaney,~A.; Makhlouf,~S. Assessment of
  electrocoagulation for the treatment of petroleum refinery wastewater.
  \emph{Journal of environmental management} \textbf{2009}, \emph{91},
  180--185\relax
\mciteBstWouldAddEndPuncttrue
\mciteSetBstMidEndSepPunct{\mcitedefaultmidpunct}
{\mcitedefaultendpunct}{\mcitedefaultseppunct}\relax
\EndOfBibitem
\bibitem[Bukhari(2008)]{bukhari2008investigation}
Bukhari,~A.~A. Investigation of the electro-coagulation treatment process for
  the removal of total suspended solids and turbidity from municipal
  wastewater. \emph{Bioresource technology} \textbf{2008}, \emph{99},
  914--921\relax
\mciteBstWouldAddEndPuncttrue
\mciteSetBstMidEndSepPunct{\mcitedefaultmidpunct}
{\mcitedefaultendpunct}{\mcitedefaultseppunct}\relax
\EndOfBibitem
\bibitem[Khandegar and Saroha(2014)Khandegar, and
  Saroha]{khandegar2014electrochemical}
Khandegar,~V.; Saroha,~A.~K. Electrochemical treatment of textile effluent
  containing Acid Red 131 dye. \emph{Journal of Hazardous, Toxic, and
  Radioactive Waste} \textbf{2014}, \emph{18}, 38--44\relax
\mciteBstWouldAddEndPuncttrue
\mciteSetBstMidEndSepPunct{\mcitedefaultmidpunct}
{\mcitedefaultendpunct}{\mcitedefaultseppunct}\relax
\EndOfBibitem
\bibitem[Guo \latin{et~al.}(2006)Guo, Zhang, Fang, and Dou]{guo2006enhanced}
Guo,~Z.-R.; Zhang,~G.; Fang,~J.; Dou,~X. Enhanced chromium recovery from
  tanning wastewater. \emph{Journal of Cleaner Production} \textbf{2006},
  \emph{14}, 75--79\relax
\mciteBstWouldAddEndPuncttrue
\mciteSetBstMidEndSepPunct{\mcitedefaultmidpunct}
{\mcitedefaultendpunct}{\mcitedefaultseppunct}\relax
\EndOfBibitem
\bibitem[El-Ashtoukhy \latin{et~al.}(2013)El-Ashtoukhy, El-Taweel, Abdelwahab,
  and Nassef]{el2013treatment}
El-Ashtoukhy,~E.; El-Taweel,~Y.; Abdelwahab,~O.; Nassef,~E. Treatment of
  petrochemical wastewater containing phenolic compounds by electrocoagulation
  using a fixed bed electrochemical reactor. \emph{Int. J. Electrochem. Sci}
  \textbf{2013}, \emph{8}, 1534--1550\relax
\mciteBstWouldAddEndPuncttrue
\mciteSetBstMidEndSepPunct{\mcitedefaultmidpunct}
{\mcitedefaultendpunct}{\mcitedefaultseppunct}\relax
\EndOfBibitem
\bibitem[Golder \latin{et~al.}(2007)Golder, Samanta, and
  Ray]{golder2007removal}
Golder,~A.~K.; Samanta,~A.~N.; Ray,~S. Removal of trivalent chromium by
  electrocoagulation. \emph{Separation and purification technology}
  \textbf{2007}, \emph{53}, 33--41\relax
\mciteBstWouldAddEndPuncttrue
\mciteSetBstMidEndSepPunct{\mcitedefaultmidpunct}
{\mcitedefaultendpunct}{\mcitedefaultseppunct}\relax
\EndOfBibitem
\bibitem[Shalaby \latin{et~al.}(2014)Shalaby, Nassef, Mubark, and
  Hussein]{shalaby2014phosphate}
Shalaby,~A.; Nassef,~E.; Mubark,~A.; Hussein,~M. Phosphate removal from
  wastewater by electrocoagulation using aluminium electrodes. \emph{American
  Journal of Environmental Engineering and Science} \textbf{2014}, \emph{1},
  90--98\relax
\mciteBstWouldAddEndPuncttrue
\mciteSetBstMidEndSepPunct{\mcitedefaultmidpunct}
{\mcitedefaultendpunct}{\mcitedefaultseppunct}\relax
\EndOfBibitem
\end{mcitethebibliography}

\end{document}